\documentclass[floatfix,10pt,onecolumn,showpacs,amsmath,amssymb]{revtex4}

\usepackage{epsf}
\usepackage{graphicx}  % Include figure files
\usepackage{dcolumn}   % Align table columns on decimal point
\usepackage{bm}        % bold math
\usepackage{color}

\newcommand{\be}{\begin{equation}}
\newcommand{\en}{\end{equation}}
 \newcommand{\bea}{\begin{eqnarray}}
 \newcommand{\ena}{\end{eqnarray}}
  \newcommand{\sch}{Schwarzschild}

\begin{document}

\title{Generalized Vaidya Solutions and Misner-Sharp mass for $n$-dimensional massive gravity}
\author{Ya-Peng Hu$^{1,2,4}$ \footnote{Electronic address: huyp@nuaa.edu.cn},Xin-Meng Wu$^{1}$ \footnote{Electronic address: wuxm@nuaa.edu.cn},
Hongsheng Zhang$^{3,4,5}$\footnote{Electronic address: hongsheng@shnu.edu.cn}}
\affiliation{ $^1$ College of Science, Nanjing University of Aeronautics and Astronautics, Nanjing 210016, China\\
$^2$ Instituut-Lorentz for Theoretical Physics, Leiden University, Niels Bohrweg 2, Leiden 2333 CA, The Netherlands \\
$^3$ School of Physics and Technology, University of Jinan, 336 West Road of Nan Xinzhuang, Jinan, Shandong 250022, China\\
$^4$ Key Laboratory of Theoretical Physics, Institute of Theoretical Physics, Chinese Academy of Sciences, Beijing, 100190, China\\
$^5$ Center for Astrophysics, Shanghai Normal University, 100 Guilin Road, Shanghai 200234, China }

%\date{ \today}

\begin{abstract}
    Dynamical solutions are always of interest to people in gravity theories. We derive a series of generalized Vaidya solutions in the $n$-dimensional de Rham-Gabadadze-Tolley (dRGT) massive gravity with a singular reference metric. Similar to the case of the Einstein gravity, the generalized Vaidya solution can describe shining/absorbing stars. Moreover, we also find a more general Vaidya-like solution by introducing a more generic matter field than the pure radiation in the original Vaidya spacetime. As a result, the above generalized Vaidya solution is naturally included in this Vaidya-like solution as a special case. We investigate the thermodynamics for this Vaidya-like spacetime by using the unified first law, and present the generalized Misner-Sharp mass. Our results show that the generalized Minser-Sharp mass does exist in this spacetime. In addition, the usual Clausius relation $\delta Q= TdS$ holds on the apparent horizon, which implicates that the massive gravity is in a thermodynamic equilibrium state. We find that the work density vanishes for the generalized Vaidya solution, while it appears in the more general Vaidya-like solution. Furthermore, the covariant generalized Minser-Sharp mass in the $n$-dimensional de Rham-Gabadadze-Tolley massive gravity is also derived by taking a general metric ansatz into account.

\end{abstract}

%\keywords{Misner-Sharp mass; f(R) gravity; conserved charge}

%\preprint{arXiv: }
 \maketitle

\section{Introduction}
Massive gravities are significant and fundamental extensions of the Einstein gravity, however, in opposite to our intuition, to endow a mass to the graviton is not an easy problem. In 1939, Fierz and Pauli first introduced the linear massive gravity theory \cite{FP}. Note that, a massless graviton has only two polarizations,  and a sound massive gravity theory generally has five degree of freedoms. However, the surplus three degree of freedoms are proved to be intractable when the mass of graviton vanishes in the linear massive gravity \cite{rew}.  To overcome this problem, one tries to introduce the non-linear massive gravities, but a more serious problem, the Boulware-Deser ghost problem, appears\cite{BD}. Recently, the so called de Rham-Gabadadze-Tolley (dRGT) massive gravity has been proposed \cite{deRham:2010ik, deRham:2010kj,deRham:2014zqa}, which is a nonlinear massive gravity theory and has been shown to be ghost-free \cite{Hassan:2011hr,Hassan:2011tf}. Note that, in the dRGT model, the reference metric is full rank. But yet, a singular reference metric is also  important \cite{Vegh:2013sk}, in which the ghost problem is investigated in \cite{Hu:2015xva,Zhang:2015nwy}. Moreover, according to the AdS/CFT correspondence~\cite{Maldacena:1997re,Gubser:1998bc,Witten:1998qj}, many clues show that the massive graviton in the bulk is related to some interesting effects of the dual field which resides on the UV boundary of an asymptotical AdS spacetime, i.e. the effects like a lattice to deduce the momentum dissipation~\cite{Vegh:2013sk,Blake:2013owa,Amoretti:2014zha,Hu:2015dnl}. Many researches about the dRGT massive gravity have been also done~\cite{Vegh:2013sk,Blake:2013owa,Adams:2014vza,Cai:2014znn,Hu:2015xva,Xu:2015rfa,Hendi:2015hoa,Cao:2015cza,Blake:2013bqa,Davison:2013jba,Davison:2013txa,
Zhang:2015nwy,Hu:2015dnl,Do:2016abo,Hendi:2015pda,Hu:2016mym,Amoretti:2014zha,Volkov:2013roa,Tasinato:2013rza,Babichev:2015xha}.

Among these researches, one interesting issue is to find out exact solutions in the dRGT massive gravity~\cite{Cai:2014znn,Hendi:2015pda,Hendi:2015hoa,Hu:2016mym,Volkov:2013roa,Tasinato:2013rza,Babichev:2015xha}. Usually we assume some symmetries of the spacetime when we seek a new solution. The translation invariance along a time-like Killing vector is one of the most important symmetry, but in some violent astrophysical processes, or when the mass of the matters surrounding the central celestial bodies are not negligible, such an assumption may be no longer reliable. However, finding an exact dynamical solution describing these realistic processes has proven to be an intricate issue.

Vaidya found an important dynamical toy model for a spherically symmetric spacetime \cite{vai},
\be
ds^2=-(1-\frac{2M(v)}{r})dv^2+2dvdr+r^2d\Omega_2^2,
\en
where $M(v)$ is the mass parameter, $d\Omega_2^2$ is the metric of the 2-sphere, and the stress tensor of the matter field is given by $T_{ab}=\mu l_al_b$. Here $l_a=(dv)_a$ in the above coordinates $(v,r,x^i)$ and $\mu$ is the energy density. This solution is well-known as the Vaidya solution. Note that, the Vaidya solution describes a spherically symmetric spacetime sourced by massless particles (not quanta of the Maxwell field which are called the pure radiations. In addition, since $M(v)$ is an undetermined function in the Vaidya metric, in principle, it can describe an arbitrary spherically symmetric energy flow from the central star. When $ M(v)$=constant it comes back to  the \sch~ spacetime, and when $M(v)=0$ it degenerates to the Minkowski spacetime. It should be emphasized that the Vaidya solution is an important solution since it encodes some essential properties of the dynamical spherically symmetric spacetimes, while keeps simple enough to handle. Therefore, in our paper the first task is to generalize the above Vaidya solution to a more general case, i.e., the exact generalized Vaidya and Vaidya-like solutions in the $n$-dimensional spacetime with maximally symmetric $(n-2)$-subspace in the dRGT massive gravity.  The metric ansatz reads,
\be
ds^{2}=-f(v,r)dv^2+2 dv dr+r^{2}\gamma _{ij}dx^{i}dx^{j},
\label{metricn1}
\en
where $\gamma _{ij}$ is the metric on a $(n-2)$-dimensional
constant curvature space ${\cal N}$ with its sectional curvature
  $k=\pm 1,0$, and the two-dimensional  ${\cal T}$ spanned by $(v,r)$ has the metric $h_{ab}$. In addition, during obtaining the generalized dynamical solutions,  we first adopt the pure radiation as the matter field. Then we extend the matter field to a more general case~\cite{Dominguez:2005rt,CCHK}, and then obtain a generalized Vaidya-like solution, in which the generalized Vaidya solution is included as a special case. For the generalized Vaidya solution in dRGT massive gravity, we find that it is consistent with the result in some previous works where the corresponding static solution has been found~\cite{Cai:2014znn}.

  As an important progress, black hole thermodynamics (more generally gravi-thermodynamics) significantly boosts our understandings of gravity theory. It is even treated as a critical probe to the quantum gravity theory. Gravi-thermodynamics is well established in stationary spacetime. For dynamical spacetimes, there is still no generally accepted theory yet. The first difficulty is that some key physical concepts, including temperature, entropy, horizon, etc, become subtle. The second difficulty is that it is hard to define a reversible process in a dynamical spacetime. However, some researches have shown that the unified first law is a nice approach in gravi-thermodynamics if the spacetime has an maximally symmetric subspace, since usually it can be directly derived from the field equation itself~\cite{Hayward,mae,CCHK,Cai-Kim,Cai:2008ys,Hu:2015xva} in such spacetimes. Thus, one can apply it in a dynamical spacetime without essential obstructions for these spacetimes. In our paper, we apply the unified first law to investigate the thermodynamics of the above generalized dynamical solutions in dRGT massive gravity. Note that, the Misner-Sharp mass is a significant quantity in the unified first law. In Einstein's general relativity, the Misner-Sharp mass always exists~\cite{Hayward,mae,CCHK,Cai-Kim,Cai:2008ys,Hu:2015xva}. In addition, since it encodes rich information of the corresponding gravity field \cite{self2}, one can obtain a series of exact solutions through thermodynamic approaches from the Misner-Sharp mass \cite{self31}. However, the generalized Misner-Sharp mass may be absent in some modified gravity like $f(R)$ gravity ~\cite{Cai:2009qf,Zhang:2014goa}.

   In our case, by using the unified first law, we find that the generalized Misner-Sharp mass does exist for the above generalized dynamical solutions, and obtain the first law of thermodynamics on the apparent horizon for these generalized dynamical solutions. In addition, the usual Clausius relation $\delta Q= TdS$ holds on the apparent horizon, which implies that the dRGT massive gravity is in a thermodynamic equilibrium state~\cite{aka,Cai:2009qf,Eling:2006aw,Hu:2015xva}.  It should be emphasized that the existence of the Misner-Sharp mass in some special solutions does not always imply the existence of it in the corresponding gravity theory. For example, the Misner-Sharp mass exists in the FRW solution and static solution in $f(R)$ gravity. However, it does not always exist in a general spherically symmetric spacetime in $f(R)$ gravity~\cite{Cai:2009qf,Zhang:2014goa}. Essentially, the generalized Misner-Sharp mass is a conserved charge of the spacetime corresponding to the Kodama vector (reduced to a Killing one in stationary spacetime), which depends on the gravity theory in consideration~\cite{mae,Cai:2009qf,Zhang:2014goa}. The integrability of such a conserved charge, and thus the existence of the generalized Misner-Sharp mass, is a non-trivial problem. Therefore, we need further study the existence of the generalized Misner-Sharp mass in a general spacetime with maximally symmetric subspaces. We show that the generalized Misner-Sharp mass in the $n$-dimensional dRGT massive gravity indeed exist, and the covariant form has also been obtained, i.e., the result is not constrained to any special solution.

This paper is organized as follows. In section II, we first obtain the generalized Vaidya solution in the dRGT massive gravity, and then consider a more general matter field to obtain a generalized Vaidya-like solution. In section III, we  use the unified first law to investigate the thermodynamics of these generalized dynamical solutions. Our results show that the generalized Misner-Sharp mass exists in these solutions. In Section IV, we further derive the covariant generalized Misner-Sharp mass for the $n$-dimensional dRGT massive gravity by considering the more general metric ansatz and matter fields. Finally, we draw the conclusions and discussions in Section V.

\section{Generalized dynamical solutions in the $N$-dimensional massive gravity}
In this section, we explore the generalized dynamical solutions in the $n$-dimensional dRGT massive gravity. The action of the dRGT massive gravity in an $n$-dimensional spacetime with a cosmological constant $\Lambda=-\frac{(n-1)(n-2)}{2\ell^2}$ reads~\cite{Vegh:2013sk,Cai:2014znn},
\begin{equation}
\label{actionmassive}
S =\frac{1}{16\pi G}\int d^{n}x \sqrt{-g} \left[ R +\frac{(n-1)(n-2)}{\ell^2} +m^2 \sum^4_i c_i {\cal U}_i (g,f)\right],
\end{equation}
where   $f$ is a constant symmetric tensor, which is usually called the  reference metric,
$c_i$ and $\ell$ are constants, and ${\cal U}_i$ are symmetric polynomials of the eigenvalues of the $n \times n$ matrix ${\cal K}^{\mu}_{\ \nu} \equiv \sqrt {g^{\mu\alpha}f_{\alpha\nu}}$~,
\begin{eqnarray}
\label{eq2}
&& {\cal U}_1= [{\cal K}], \nonumber \\
&& {\cal U}_2=  [{\cal K}]^2 -[{\cal K}^2], \nonumber \\
&& {\cal U}_3= [{\cal K}]^3 - 3[{\cal K}][{\cal K}^2]+ 2[{\cal K}^3], \nonumber \\
&& {\cal U}_4= [{\cal K}]^4- 6[{\cal K}^2][{\cal K}]^2 + 8[{\cal K}^3][{\cal K}]+3[{\cal K}^2]^2 -6[{\cal K}^4].
\end{eqnarray}
The square root in ${\cal K}$ means $(\sqrt{A})^{\mu}_{\ \nu}(\sqrt{A})^{\nu}_{\ \lambda}=A^{\mu}_{\ \lambda}$ and $[{\cal K}]=K^{\mu}_{\ \mu}=\sqrt {g^{\mu\alpha}f_{\alpha\mu}}$.

From the action and considering the matter fields, the equations of motion are
\begin{eqnarray}
\mathcal{G}_{\mu\nu}\equiv R_{\mu\nu}-\frac{1}{2}Rg_{\mu\nu}-\frac{(n-1)(n-2)}{2\ell^2} g_{\mu\nu}+m^2 \chi_{\mu\nu}&=&8\pi G T_{\mu \nu },~~
\label{fd1}
\end{eqnarray}
where
\begin{eqnarray}
&& \chi_{\mu\nu}=-\frac{c_1}{2}({\cal U}_1g_{\mu\nu}-{\cal K}_{\mu\nu})-\frac{c_2}{2}({\cal U}_2g_{\mu\nu}-2{\cal U}_1{\cal K}_{\mu\nu}+2{\cal K}^2_{\mu\nu})
-\frac{c_3}{2}({\cal U}_3g_{\mu\nu}-3{\cal U}_2{\cal K}_{\mu\nu}\nonumber \\
&&~~~~~~~~~ +6{\cal U}_1{\cal K}^2_{\mu\nu}-6{\cal K}^3_{\mu\nu})
-\frac{c_4}{2}({\cal U}_4g_{\mu\nu}-4{\cal U}_3{\cal K}_{\mu\nu}+12{\cal U}_2{\cal K}^2_{\mu\nu}-24{\cal U}_1{\cal K}^3_{\mu\nu}+24{\cal K}^4_{\mu\nu}).
\end{eqnarray}

In this article, we will investigate the generalized dynamical solutions in the $n$-dimensional spacetime with a maximally symmetric inner space in the dRGT massive gravity, and the metric ansatz is just (\ref{metricn1}). For this metric ansatz, we take the following reference metric as in Ref. \cite{Cai:2014znn}
\begin{equation}
\label{reference}
f_{\mu\nu} = {\rm diag}(0,0, c_0^2 \gamma_{ij} ),
\end{equation}
with $c_0$ is a positive constant. Thus,
\begin{eqnarray}
[{\cal K}]=\frac{n-2}{r}c_0,~[{\cal K}^2]=\frac{n-2}{r^2}c_0^2,~[{\cal K}^3]=\frac{n-2}{r^3}c_0^3,~[{\cal K}^4]=\frac{n-2}{r^4}c_0^4
\end{eqnarray}
with the symmetric polynomials become
\begin{eqnarray}
&&{\cal U}_1= \frac{(n-2)c_0}{r}, \\
&&{\cal U}_2= \frac{(n-2)(n-3)c_0^2}{r^2}, \\
&&{\cal U}_3= \frac{(n-2)(n-3)(n-4)c_0^3}{r^3}, \\
&&{\cal U}_4= \frac{(n-2)(n-3)(n-4)(n-5)c_0^4}{r^4},
\end{eqnarray}
and the corresponding components of $\mathcal{G}_{\mu\nu}$ are
\begin{eqnarray}
\mathcal{G}_{v}^{v}&=&\mathcal{G}_{r}^{r}=\Lambda+\frac{n-2}{2}\times [\frac{(r^{n-3}f)'-(n-3)r^{n-4}k-c_1c_0m^2r^{n-3}-(n-3)c_2c_{0}^2m^2r^{n-4}-(n-3)(n-4)c_3c_{0}^3m^2r^{n-5}}{r^{n-2}} \nonumber\\
&-&\frac{(n-3)(n-4)(n-5)c_4c_{0}^4m^2r^{n-6}}{r^{n-2}}], \label{MainEq}\\
\mathcal{G}^{i}_{j}&=&\delta^{i}_{j}\times [\Lambda+\frac{(r^{n-3}f)''-(n-3)(n-4)r^{n-5}k-(n-3)c_1c_0m^2r^{n-4}-(n-3)(n-4)c_2c_{0}^2m^2r^{n-5}}{2r^{n-3}}\nonumber\\
&-&\frac{(n-3)(n-4)(n-5)c_3c_{0}^3m^2r^{n-6}-(n-3)(n-4)(n-5)(n-6)c_4c_{0}^4m^2r^{n-7}}{2r^{n-3}}],\\
\mathcal{G}^{r}_{v}&=&\frac{-(n-2)\dot{f}}{2r},\label{EqST}\\
\mathcal{G}^{v}_{r}&=&0.
\end{eqnarray}
where a prime/overdot denotes the derivative with respect to $r/v$. In the followings, we investigate two cases by considering different matter fields. In the first case, the generalized Vaidya solution is derived sourced by the pure radiations in analogy to the Vaidya solution in the Einstein gravity. In the second case, we consider a more generic source matter than the usual pure radiations to obtain a more general dynamical solution, i.e., a generalized Vaidya-like solution.

\subsection{Special case: Generalized Vaidya Solution}

For the pure radiations, the stress energy tensor is given by $T_{ab}=\mu l_al_b$, where $l_a=(dv)_a$ is expressed in the coordinates $(v,r,x^i)$ in (\ref{metricn1}). The components of the field equation (\ref{fd1}) corresponding to the metric (\ref{metricn1}) present,
\begin{eqnarray}
&&\mathcal{G}_{v}^{v}=\mathcal{G}_{r}^{r}=\Lambda+\frac{n-2}{2}\times [\frac{(r^{n-3}f)'-(n-3)r^{n-4}k-c_1c_0m^2r^{n-3}-(n-3)c_2c_{0}^2m^2r^{n-4}-(n-3)(n-4)c_3c_{0}^3m^2r^{n-5}}{r^{n-2}} \nonumber\\
&&-\frac{(n-3)(n-4)(n-5)c_4c_{0}^4m^2r^{n-6}}{r^{n-2}}]=0, \label{MainEq1}\\
&&\mathcal{G}^{i}_{j}=\delta^{i}_{j}\times [\Lambda+\frac{(r^{n-3}f)''-(n-3)(n-4)r^{n-5}k-(n-3)c_1c_0m^2r^{n-4}-(n-3)(n-4)c_2c_{0}^2m^2r^{n-5}}{2r^{n-3}}\nonumber\\
&&-\frac{(n-3)(n-4)(n-5)c_3c_{0}^3m^2r^{n-6}-(n-3)(n-4)(n-5)(n-6)c_4c_{0}^4m^2r^{n-7}}{2r^{n-3}}]=0,\\
&&\mathcal{G}^{r}_{v}=\frac{-(n-2)\dot{f}}{2r}=8\pi G \mu,\label{EqST1}\\
&&\mathcal{G}^{v}_{r}=0,
\end{eqnarray}

Note that the components $\mathcal{G}^{i}_{j}$ are not independent, because they are linear combination of the terms of $\mathcal{G}_{v}^{v}$ and $\partial_r{\mathcal{G}_{v}^{v}}$,
\begin{eqnarray}
\mathcal{G}^{i}_{j}=\delta^{i}_{j}[\mathcal{G}_{v}^{v}+r\partial_r{\mathcal{G}_{v}^{v}}/(n-2)]=\delta^{i}_{j}[\frac{1}{(n-2)r^{n-3}}\partial_r(r^{n-2}\mathcal{G}_{v}^{v})].
\end{eqnarray}
Therefore, $\mathcal{G}^{i}_{j}=0$ do not yield independent equations. From the above equation in (\ref{MainEq1}), we can easily obtain the generalized Vaidya solution in the $n$-dimensional dRGT massive gravity,
\begin{eqnarray}
&&f(v,r)=k+\frac{r^2}{\ell^2}-\frac{M(v)}{r^{n-3}}+\frac{c_0c_1m^2}{n-2}r+c_0^2c_2m^2+\frac{(n-3)c_0^3c_3m^2}{r}+\frac{(n-3)(n-4)c_0^4c_4m^2}{r^2},\label{GVAdSSolution}
\end{eqnarray}
with
\begin{eqnarray}
&&\mu=-\frac{(n-2)\dot{f}}{16\pi G r}=\frac{(n-2)\dot{M}(v)}{16\pi G r^{n-2}}, \label{solution}
\end{eqnarray}
which can be  obtained by inserting (\ref{GVAdSSolution}) into (\ref{EqST1}), and $M(v)$ is the mass parameter. Our solution is consistent with the result in some previous work like \cite{Cai:2014znn}. Since if $M(v)$ is independent of $v$, i.e. a constant, and hence $f(v,r)$ can be written as $f(r)$, then after the transformation in the metric ansatz (\ref{metricn1})
\begin{eqnarray}
dv=dt+\frac{1}{f(r)}dr,
\end{eqnarray}
the above solution (\ref{GVAdSSolution}) comes back to the static solution in $n$-dimensional spacetime found in \cite{Cai:2014znn}.

\subsection{The general case: Generalized Vaidya-like Solution}

Now we further generalize the above generalized Vaidya solution in the dRGT massive gravity
to a more general case. For a general discussion of the stress energy form constrained by the energy conditions in the Vaidya-type solutions, see~\cite{Wang:1998qx}.
%in analogy to the Vaidya solution in Lovelock gravity\cite{self5}
Note that, for the metric
(\ref{metricn1}) and the reference metric (\ref{reference}), we have $\mathcal{G}^r_r=\mathcal{G}^v_v$, so
the energy-momentum tensor of matter field should satisfy $T^r_r=T^v_v$.
Certainly, the pure radiation matter discussed above
satisfies the constraint. In fact, they are $T^r_r=T^v_v=0$. Therefore, if we relax this condition to $T^{i}_{i}=\sigma T^r_r=\sigma T^v_v$ (where $\sigma$ is a
constant, and the equation does not sum over $i$), then from the
equation $\nabla_{\mu}T^{\mu}_{\nu}=0$ or the explicit expressions of
$\mathcal{G}^{\mu}_{\nu}$ in equations (\ref{MainEq}) to (\ref{EqST}), we can derive
\begin{equation}
\label{TRVEq}
\partial_v T^v_v+\partial_rT^r_v+\frac{n-2}{r}T^r_v=0\, ,
\end{equation}
and
\begin{equation}
\label{TRREq}
\partial_r T^v_v+\frac{(n-2)(1-\sigma)}{r}T^v_v=0\, .
\end{equation}
So, for the pure radiation matter with $T^r_r=T^v_v=0$, one finds that $T^r_v$ has to be proportional to $1/r^{n-2}$, which is consistent with the above generalized Vaidya case in (\ref{solution}).

Therefore, for the more general case $T^r_r=T^v_v\ne 0$ for the matter field, and hence from the equation (\ref{TRREq}), $T^r_r$ and $T^v_v$ should satisfy
\begin{equation}
T^r_r=T^v_v=\mathcal{C}(v) r^{-(n-2)(1-\sigma)}\, ,
\end{equation}
where $\mathcal{C}(v)$ is a function of $v$. In addition, the off-diagonal part
of the energy-momentum tensor $T^{\mu}_{\nu}$, i.e., the component $T^r_v$
has to satisfy the equation (\ref{TRVEq}). Now the equation (\ref{MainEq1}) is modified as
\begin{equation}
\mathcal{G}^v_v=8\pi G~\mathcal{C}(v) r^{-(n-2)(1-\sigma)}\,.
\label{Fequationgeneral}
\end{equation}
Integrating this equation, we obtain the expression of $f(v,r)$
\begin{eqnarray}
f(v,r)&=&k+\frac{r^2}{\ell^2}+\frac{c_0c_1m^2}{n-2}r+c_0^2c_2m^2+\frac{(n-3)c_0^3c_3m^2}{r}+\frac{(n-3)(n-4)c_0^4c_4m^2}{r^2}\notag \\
&-&\frac{M(v)}{r^{n-3}}+\frac{16\pi G}{(n-2)r^{n-3}}\mathcal{C}(v)\Theta (r), \label{Solution2}
\end{eqnarray}
where $M(v)$ is an arbitrary function of $v$, and $\Theta(r) = \int dr r^{(n-2)\sigma}$. In details, when $\sigma=-1/(n-2)$,
\begin{equation}
\Theta(r)= \mathrm{ln}
(r)\, ,
\end{equation}
and in other cases
\begin{equation}
\Theta(r)= \frac{r^{(n-2)\sigma+1}}{(n-2)\sigma+1}.
\end{equation}
Note that the parameter $\sigma$ and functions $m(v)$
and $\mathcal{C}(v)$ should satisfy some consistency relations if one imposes some energy condition for the
energy-momentum tensor. In addition, from (\ref{EqST}), we have
\begin{equation}
\label{mutilde} T^r_v=\tilde{\mu}=\frac{(n-2)\dot{M}(v)}{16\pi G
r^{n-2}}-\frac{\dot{\mathcal{C}}(v)\Theta (r)}{r^{n-2}}\, .
\end{equation}
which is also consistent with the equation (\ref{TRVEq}). Therefore, we have also obtained the stress tensor of matter field in this more general case. More precisely, we can further write the stress tensor of matter field in this more general case as
\begin{equation}
\label{energymomentumtensor} T_{ab}=\tilde{\mu}
l_al_b-P(l_an_b+n_al_b)+\sigma P q_{ab}\, ,
\end{equation}
where $n_a$ is a null vector which satisfies $l_an^a=-1$. In
coordinates $(v,r,x^i)$, $l_a=(dv)_a$ and
$n_a=f/2(dv)_a-(dr)_a$, while the tensor $q_{ab}$ is a projection
operator given by $q_{ab}=g_{ab}+l_an_b+l_bn_a$, and the quantity $P$ is the radial pressure with
the form $P=\mathcal{C}(v)r^{-(n-2)(1-\sigma)}$. In addition, the
metric (\ref{metricn1}) can be put into the form
$g_{ab}=h_{ab}+q_{ab}$, where
\begin{equation}
h_{ab}=-l_an_b-l_bn_a
\end{equation}
is the metric of two-dimensional spacetime ${\cal T}$ spanned by the coordinates $(v,r)$. Certainly, in the coordinates
$(v,r,x^i)$, the line element of $h_{ab}$ can be expressed as $-f(v,r)dv^2+2dvdr$. Therefore, (\ref{Solution2}) together with (\ref{energymomentumtensor}) is  a more general case with new dynamical solution, which we call the generalized Vaidya-like solution. Obviously, the above generalized Vaidya solution is  a special case of this generalized Vaidya-like solution with $\mathcal{C}(v)=0$.

\section{Thermodynamics of the generalized dynamical solutions}
In this section, we will investigate thermodynamics of the above generalized dynamical solutions in the dRGT massive gravity by using the unified first law, and we concentrate on the generalized Vaidya-like solution obtained in the more general case, since it naturally includes the generalized Vaidya solution as a special case. According to the unified first law, similar to
the case of the Einstein gravity \cite{Hayward}, one can formally cast the
equation (\ref{fd1}) of gravitational field into the form,
\begin{equation}
dM_{eff}=A\Psi_a dx^a +WdV, \label{2eq5}
\end{equation}
where $A=V_{k}r^{n-2}$ and $V =V_{k}r^{n-1}/(n-1)$ are the
area and volume of the $(n-2)$-dimensional constant curvature space ${\cal N}$ with radius $r$, $W$ is called work density defined
as $W= -h^{ab}T_{ab}/2$ and $\Psi_a$ is the energy supply vector with the definition
$\Psi_a=T_a^{\ b} \partial _b r +W \partial_a r$. Here, $T_{ab}$ is the projection of the stress tensor
$T_{\mu\nu}$ of matter into $h_{ab}$.

After substituting the explicit forms of generalized dynamical solutions in the dRGT massive gravity (\ref{Solution2})  and (\ref{energymomentumtensor}), we can explicitly obtain the following quantities,
\begin{eqnarray}
&&W=-P,~~ \Psi_a=\tilde{\mu} l_a,\\
&&A\Psi_a dx^a +WdV=V_kr^{n-2}\tilde{\mu}dv-PV_kr^{n-2}dr\equiv X(v,r)dv+Y(v,r)dr.
\end{eqnarray}
It is easy to check,
       \be
       \frac{\partial X(v,r)}{\partial r}=\frac{\partial Y(v,r)}{\partial v},
       \en
            which ensures that $dM_{eff}$ is a closed form, and thus qualified as the generalized Misner-Sharp mass for the above generalized dynamical solutions in the dRGT massive gravity. Moreover, the generalized Misner-Sharp mass can be easily obtained  in this case,
\begin{eqnarray}
M_{eff}=V_k[\frac{(n-2)M(v)}{16 \pi G }-\mathcal{C}(v)\Theta (r)], \label{Meff1}
\end{eqnarray}

Next, we will use the unified first law and generalized Misner-Sharp mass (\ref{Meff1}) to investigate the thermodynamics of the above generalized dynamical solutions on the apparent horizon $r_A$, where $r_A$ is defined as the trapped surface $h^{ab}\partial_a{r}\partial_b{r}=0$. In our case, we can easily obtain the location of the apparent horizon $r_A$ is $f(v,r)=0$ in (\ref{Solution2}). On the apparent horizon, the energy flow across the
apparent horizon is~\cite{CCHK,Cai-Kim,Cai:2008ys,Hu:2015xva}
\begin{equation}
\delta Q=dM_{eff}|_{r_{A}}=A\Psi_adx^a|_{r=r_A} = A\Psi_v dv=-\frac{(n-2)V_kr_A^{n-3}}{16\pi G}\dot{f}(r_A)dv.
\end{equation}
On the other hand, the temperature of generalized dynamical solution is $T=\frac{\kappa}{2\pi}$, where the surface gravity $\kappa$ defined on the apparent horizon is $\kappa=D_aD^ar=\frac{1}{2\sqrt{-h}}\frac{\partial}{\partial x^{\mu}}(\sqrt{-h}h^{\mu\nu}\partial_v r)=f'(r_A)/2$~\cite{Hayward,mae,CCHK,Cai-Kim,Cai:2008ys,Hu:2015xva}. Here, $D_a$ is the covariant derivative associated with metric $h_{ab}$. In addition, the entropy of apparent horizon is $S=\frac{A}{4G}=\frac{V_kr_A^{n-2}}{4G}$~\cite{Cai:2014znn}. Therefore,
\begin{equation}
TdS=\frac{\kappa}{2\pi}dS=\frac{(n-2)V_kr_A^{n-3}}{16\pi G}f'(r_A)\dot{r}_Adv. \label{Clausius}
\end{equation}
After using the simple relation $f'(r_A)\dot{r}_A=-\dot{f}(r_A)$ derived from $f(r_A,v)=0$, we can easily obtain that the usual Clausius relation
$\delta Q= TdS$ does hold on the apparent horizon of the generalized dynamical solution, which indicates that the dRGT massive gravity is an equilibrium state~\cite{Eling:2006aw}. Note that, this result is consistent with the investigation in \cite{Hu:2015xva} by taking the FRW universe into account. In addition, it should be emphasized that the usual Clausius relation $\delta Q= TdS$ does not always hold on the apparent horizon. For example, the usual Clausius relation does not hold for the $f(R)$ gravity, which can be treated as the effects of the nonequilibrium of the space-time~\cite{aka,Cai:2009qf,Eling:2006aw}. Therefore, after taking (\ref{Clausius}) and Clausius relation into account, the unified first law in (\ref{2eq5}) on the apparent can be rewritten as
\begin{equation}
dM_{\rm eff}=TdS +WdV, \label{FirstLaw}
\end{equation}
which is just the first law of thermodynamics for the generalized Vaidya-like solution. Note that, the work density $W$ in (\ref{FirstLaw}) is  nonzero for the generalized Vaidya-like solution, which makes another difference from the generalized Vaidya solution whose $W=0$.

\section{Generalized Misner-Sharp mass for the $N$-dimensional massive gravity}
Note that, the Misner-Sharp mass is a quantity depending on not only the symmetry in the solution, i.e. usually just defined in a spacetime with a maximally symmetric subspace, but also the underlying gravity theory. Hence, the existence of Minser-Sharp mass in a special solution with a maximally symmetric subspace does not always guarantee its existence in the gravity for the general solutions with the same maximally symmetric subspace, for example the $f(R)$ gravity\cite{Cai:2009qf,Zhang:2014goa}. Therefore, we should further investigate the existence of the Misner-Sharp mass in a general spacetime with a maximally symmetric subspace. Moreover, in order to investigate the generalized Misner-Sharp mass for the $n$-dimensional dRGT massive gravity, we usually write down the more general metric ansatz in a double-null coordinates as follows,
\bea
ds^{2}=-2e^{-\varphi(u,v)}dudv+r^{2}(u,v)\gamma _{ij}dx^{i}dx^{j}.
\label{metricn}
\ena
 here $\gamma _{ij}$ is the metric on the maximally symmetric  subspace same as in (\ref{metricn1}).
In the coordinates (\ref{metricn}) the RHS in (\ref{2eq5}) reads,
\be
A\Psi_a dx^a +WdV=A(u,v)du+B(u,v)dv,
 \label{AB}
\en
where
\be
\label{AB1}
A(u,v)=V_{k}r^{n-2}e^{\varphi}(r,_{u}T_{uv}-r,_{v}T_{uu}),
\en
\be
B(u,v)
=V_{k}r^{n-2}e^{\varphi}(r,_{v}T_{uv}-r,_{u}T_{vv}).
\label{AB2}
\en
 Here a comma denotes partial derivative. Substituting (\ref{2eq5}) into (\ref{AB}), we reach
 \be
 F\equiv dM_{eff}=A(u,v)du+B(u,v)dv.
 \label{F}
 \en
The components of the field equation (\ref{fd1}) in the coordinates (\ref{metricn}) read,
\bea
8\pi GT_{uu}&=& -(n-2)\frac{\varphi,_{u}r,_{u}+r,_{uu}}{r},  \notag \\
8\pi G T_{vv}&=& -(n-2)\frac{\varphi,_{v}r,_{v}+r,_{vv}}{r},  \notag \\
8\pi G T_{uv}&=&\frac{-\Lambda}{e^{\varphi}}+\frac{n-2}{2e^{\varphi}r}(2r,_{uv}e^{\varphi}+c_1c_0m^2) +\frac{(n-2)(n-3)(k+2e^{\varphi}r,_{u}r,_{v}+c_2c_0^2m^2)}{2e^{\varphi}r^2}\notag \\
&&+\frac{(n-2)(n-3)(n-4)c_3c_0^3m^2}{2e^{\varphi}r^3}+\frac{(n-2)(n-3)(n-4)(n-5)c_4c_0^4m^2}{2e^{\varphi}r^4}.
\label{Tcompt}
\ena
Obviously, a well-defined $M_{eff}$ in (\ref{F}) requires $F$ is a closed form $dF=0$, which means,
    \be
    A_{,v}~ dv\wedge du+B_{,u}~ du\wedge dv=0.
    \en
    Then we obtain the constraint for a well-defined $M_{eff}$,
     \be
     A_{,v}=B_{,u}.
     \label{AvBu}
     \en
 Substituting (\ref{Tcompt}) into (\ref{AB1}) and (\ref{AB2}), we  obtain
\begin{eqnarray}
A(u,v)&=&\frac{V_k}{8\pi G} [-\Lambda r,_{u}r^{n-2}+(n-2)r^{n-3}e^{\varphi}r,_{u}r,_{vu}+\frac{k}{2}(n-2)(n-3)r^{n-4}r,_{u}+(n-2)(n-3)e^{\varphi}r,_{v}r,_{u}^2r^{n-4}\notag\\
&+&e^{\varphi}r^{n-3}(n-2)(r,_{u}r,_{v}+r,_{v}r,_{uv})+\frac{(n-2)r^{n-3}r,_{u}c_1c_0m^2}{2}+\frac{(n-2)(n-3)r^{n-4}r,_{u}c_2c_0^2m^2}{2}\notag\\
&+& \frac{(n-2)(n-3)(n-4)r^{n-5}r,_{u}c_3c_0^3m^2}{2}+\frac{(n-2)(n-3)(n-4)(n-5)r^{n-6}r,_{u}c_4c_0^4m^2}{2}],\notag\\
B(u,v)&=&\frac{V_k}{8\pi G} [-\Lambda r,_{v}r^{n-2}+(n-2)r^{n-3}e^{\varphi}r,_{v}r,_{vu}+\frac{k}{2}(n-2)(n-3)r^{n-4}r,_{v}+(n-2)(n-3)e^{\varphi}r,_{u}r,_{v}^2r^{n-4}\notag\\
&+&e^{\varphi}r^{n-3}(n-2)(r,_{v}r,_{u}+r,_{u}r,_{uv})+\frac{(n-2)r^{n-3}r,_{v}c_1c_0m^2}{2}+\frac{(n-2)(n-3)r^{n-4}r,_{v}c_2c_0^2m^2}{2}\notag\\
&+& \frac{(n-2)(n-3)(n-4)r^{n-5}r,_{v}c_3c_0^3m^2}{2}+\frac{(n-2)(n-3)(n-4)(n-5)r^{n-6}r,_{v}c_4c_0^4m^2}{2}].\label{EPA}
\end{eqnarray}
Using the above explicit forms of $A(u,v)$ and $B(u,v)$ , we find that the above constraint is automatically  satisfied for the $n$-dimensional dRGT massive gravity, which guarantees that $M_{eff}$ is well-defined. Thus directly
integrating (\ref{2eq5}) presents the generalized Misner-Sharp
 mass in the $n$-dimensional dRGT massive gravity
\begin{eqnarray}
M_{eff}&=&\int A(u,v)du +\int\Big[B(u,v)-\frac{\partial}{\partial v}\int A(u,v)du
 \Big]dv \notag \\
&=&\frac{V_k(n-2)}{16 \pi G}r^{n-3}[\frac{r^2}{\ell^2}+k+2e^{\varphi}r,_{u}r,_{v}+\frac{c_0c_1m^2}{n-2}r+c_0^2c_2m^2+\frac{(n-3)c_0^3c_3m^2}{r}+\frac{(n-3)(n-4)c_0^4c_4m^2}{r^2}].\label{Meff} \end{eqnarray}
Note that, here the second term in the first line of (\ref{Meff})
in fact vanishes, and we have fixed an integration constant so that
$M_{eff}$ reduces to the Misner-Sharp mass in the Einstein gravity
when the graviton mass parameter $m$ goes to zero. Furthermore, the above generalized Misner-Sharp mass
can be rewritten in a covariant form as
\begin{eqnarray}
M_{eff}=\frac{V_k(n-2)}{16 \pi G}r^{n-3}[(k-h^{ab}\partial_a r\partial_b r)+\frac{r^2}{\ell^2}+\frac{c_0c_1m^2}{n-2}r+c_0^2c_2m^2+\frac{(n-3)c_0^3c_3m^2}{r}+\frac{(n-3)(n-4)c_0^4c_4m^2}{r^2}]\label{MS}
\end{eqnarray}
For the special case in the above generalized dynamical solution (\ref{Solution2}), one can check that the result in (\ref{Meff1}) is consistent with the generalized Misner-Sharp mass in (\ref{MS}). And (\ref{MS}) is the general definition of the generalized Misner-Sharp mass in the $n$-dimensional spacetime with maximally symmetric subspace in the dRGT massive gravity.

\section{conclusion and discussion}
In this paper, through considering the pure radiation and a more general case as the matter fields, we  obtain the generalized dynamical solutions in the $n$-dimensional dRGT massive gravity, which naturally includes the generalized Vaidya solution. By using the unified first law and the Misner-Sharp mass, we investigate thermodynamics for these solutions. Besides obtaining the first law of thermodynamics for these generalized dynamical solutions on the apparent horizon, we also check that the generalized Misner-Sharp mass exists for them. Generally, a solution has a much higher symmetry than the theory itself. The existence of the Misner-Sharp mass in a special solution does not imply the existence of it in the general case. For example, the Misner-Sharp mass exists
in the FRW solutions and static solutions in $f(R)$ gravity. However, it does not always exist in a general spherically symmetric spacetime in $f(R)$ gravity. In view of this situation, we further investigate the generalized Misner-Sharp by taking the general metric ansatz and matter field into account, and find that the generalized Misner-Sharp mass really exists in a covariant form.

Note that, in the massive gravity theory, a reference metric is required. However, the theory itself does not determine the concrete form of the reference metric. This uncertainty makes the theory become arbitrary in some degree, while delivers extra conveniences in some cases. For example, there is no \sch~ solution in the unitary gauge (Minkowskian reference metric), and thus to match the tests in the solar system a chameleon mechanism is necessary. Recently, Li et al find that the \sch~solution can be obtained if one gives up the unitary gauge \cite{LLX}. Other solutions have also been found by choosing different reference metric, for example the rotating black hole solution in the dRGT massive gravity~\cite{Babichev:2014tfa}. Therefore, it is an interesting issue to find other solutions in the dRGT massive gravity by considering different reference metrics.

In addition, it is found recently  that the dynamics of black holes and black branes are greatly simplified in the limit of a large number of spacetime dimensions N~\cite{Emparan:2015gva}. Therefore, more properties for the balck holes and black branes in the large N limit will also be an interesting issue to further investigate. Furthermore, according to the AdS/CFT correspondence, the Vaidya dynamical black branes in (\ref{GVAdSSolution}) can be related to the thermalization processes of the strongly coupled fields~\cite{Balasubramanian1, Balasubramanian2}, i.e., thermalization processes of the quark-gluon plasma (QGP) produced in ultra relativistic heavy-ion collisions at the Relativistic Heavy Ion Collider (RHIC) and the Large Hadron Collider (LHC). Therefore, the underlying dual physics of our Vaidya-like dynamical black brane in (\ref{Solution2}) is also an interesting issue to be explored further.

\section{Acknowledgments}
This work is supported by the National Natural Science Foundation of China (NSFC) (Grant Nos.11575083, 11565017, 11105004, 11075106, 11275128), the Program for Professor of Special Appointment (Eastern Scholar) at Shanghai Institutions of Higher Learning, National Education Foundation of China under grant No. 200931271104, the Fundamental Research Funds for the Central Universities (Grant No. NS2015073), and the Open Project Program of State Key Laboratory of Theoretical Physics, Institute of Theoretical Physics, Chinese Academy of Sciences, China (Grant  No. Y5KF161CJ1). In addition, Y.P Hu thanks a lot for the support from the Sino-Dutch scholarship program under the China Scholarship Council (CSC).


\begin{thebibliography}{99}



  \bibitem{FP}
 M. Fierz and W. Pauli,
 %¡°On relativistic wave equations for particles of arbitrary spin
 %in an electromagnetic ?eld,¡±
 Proc. Roy. Soc. Lond. A173, 211 (1939).

 \bibitem{rew}
  K.~Hinterbichler,
  %``Theoretical Aspects of Massive Gravity,''
  Rev.\ Mod.\ Phys.\  {\bf 84}, 671 (2012)
  doi:10.1103/RevModPhys.84.671
  [arXiv:1105.3735 [hep-th]].

\bibitem{BD}
D. G. Boulware and S. Deser, Phys. Rev. D 6, 3368 (1972).

%\cite{deRham:2014zqa}
\bibitem{deRham:2014zqa}
  C.~de Rham,
  %``Massive Gravity,''
  Living Rev.\ Rel.\  {\bf 17}, 7 (2014)
  %doi:10.12942/lrr-2014-7
  [arXiv:1401.4173 [hep-th]].
  %%CITATION = doi:10.12942/lrr-2014-7;%%
  %296 citations counted in INSPIRE as of 25 Oct 2016



%\cite{deRham:2010ik}
\bibitem{deRham:2010ik}
  C.~de Rham and G.~Gabadadze,
  %``Generalization of the Fierz-Pauli Action,''
  Phys.\ Rev.\ D {\bf 82}, 044020 (2010)
  %doi:10.1103/PhysRevD.82.044020
  [arXiv:1007.0443 [hep-th]].
  %%CITATION = doi:10.1103/PhysRevD.82.044020;%%
  %605 citations counted in INSPIRE as of 25 Oct 2016

%\cite{deRham:2010kj}
\bibitem{deRham:2010kj}
  C.~de Rham, G.~Gabadadze and A.~J.~Tolley,
  %``Resummation of Massive Gravity,''
  Phys.\ Rev.\ Lett.\  {\bf 106}, 231101 (2011)
  %doi:10.1103/PhysRevLett.106.231101
  [arXiv:1011.1232 [hep-th]].
  %%CITATION = doi:10.1103/PhysRevLett.106.231101;%%
  %776 citations counted in INSPIRE as of 25 Oct 2016



%\cite{Hassan:2011hr}
\bibitem{Hassan:2011hr}
  S.~F.~Hassan and R.~A.~Rosen,
  %``Resolving the Ghost Problem in non-Linear Massive Gravity,''
  Phys.\ Rev.\ Lett.\  {\bf 108}, 041101 (2012)
  %doi:10.1103/PhysRevLett.108.041101
  [arXiv:1106.3344 [hep-th]].
  %%CITATION = doi:10.1103/PhysRevLett.108.041101;%%
  %444 citations counted in INSPIRE as of 25 Oct 2016


%\cite{Hassan:2011tf}
\bibitem{Hassan:2011tf}
  S.~F.~Hassan, R.~A.~Rosen and A.~Schmidt-May,
  %``Ghost-free Massive Gravity with a General Reference Metric,''
  JHEP {\bf 1202}, 026 (2012)
  %doi:10.1007/JHEP02(2012)026
  [arXiv:1109.3230 [hep-th]].
  %%CITATION = doi:10.1007/JHEP02(2012)026;%%
  %258 citations counted in INSPIRE as of 25 Oct 2016

%\cite{Vegh:2013sk}
\bibitem{Vegh:2013sk}
  D.~Vegh,
  %``Holography without translational symmetry,''
  {arXiv:1301.0537 [hep-th]}.
  %%CITATION = ARXIV:1301.0537;%%
  %83 citations counted in INSPIRE as of 23 Aug 201

%%%%%%%%%%%%%%%%%%%%%%%%%%%%%%%%%%%%%%%%%%%%%%%%%%%%%%%%%%%%%%%%%%ghost problem in singular reference metric
%\cite{Hu:2015xva}
\bibitem{Hu:2015xva}
  Y.~P.~Hu and H.~Zhang,
  %``Misner-Sharp Mass and the Unified First Law in Massive Gravity,''
  Phys.\ Rev.\ D {\bf 92}, no. 2, 024006 (2015)
  %doi:10.1103/PhysRevD.92.024006
  [arXiv:1502.00069 [hep-th]].
  %%CITATION = doi:10.1103/PhysRevD.92.024006;%%
  %7 citations counted in INSPIRE as of 23 Nov 2016


%\cite{Zhang:2015nwy}
\bibitem{Zhang:2015nwy}
  H.~Zhang and X.~Z.~Li,
  %``Ghost free massive gravity with singular reference metrics,''
  Phys.\ Rev.\ D {\bf 93}, no. 12, 124039 (2016)
  %doi:10.1103/PhysRevD.93.124039
  [arXiv:1510.03204 [gr-qc]].
  %%CITATION = doi:10.1103/PhysRevD.93.124039;%%
  %4 citations counted in INSPIRE as of 31 Oct 2016

%%%%%%%%%%%%%%%%%%%%%%%%%%%%%%%%%%%%%%%%%%%%%%%%%%%%%%%%%%%%%%%%%%%%%%%%%%%%%%%%%%AdS/CFT correspondence
%\begin{thebibliography}{99}
\baselineskip 12pt
%\cite{Maldacena:1997re}
\bibitem{Maldacena:1997re}
  J.~M.~Maldacena,
  %``The large N limit of superconformal field theories and supergravity,''
  Adv.\ Theor.\ Math.\ Phys.\  {\bf 2}, 231 (1998)
  [Int.\ J.\ Theor.\ Phys.\  {\bf 38}, 1113 (1999)]
  [arXiv:hep-th/9711200].
  %%CITATION = IJTPB,38,1113;%%
%\cite{Gubser:1998bc}
\bibitem{Gubser:1998bc}
  S.~S.~Gubser, I.~R.~Klebanov and A.~M.~Polyakov,
  %``Gauge theory correlators from non-critical string theory,''
  Phys.\ Lett.\  B {\bf 428}, 105 (1998)
  [arXiv:hep-th/9802109].
  %%CITATION = PHLTA,B428,105;%%
%\cite{Witten:1998qj}
\bibitem{Witten:1998qj}
  E.~Witten,
  %``Anti-de Sitter space and holography,''
  Adv.\ Theor.\ Math.\ Phys.\  {\bf 2}, 253 (1998)
  [arXiv:hep-th/9802150].
  %%CITATION = 00203,2,253;%%
%\cite{Aharony:1999ti}

%\bibitem{Aharony:1999ti}O.~Aharony, S.~S.~Gubser, J.~M.~Maldacena, H.~Ooguri and Y.~Oz,
  %``Large N field theories, string theory and gravity,''Phys.\ Rept.\  {\bf 323}, 183 (2000) [arXiv:hep-th/9905111].
  %%CITATION = PRPLC,323,183;%%





%%%%%%%%%%%%%%%%%%%%%%%%%%%%%%%%%%%%%%%%%%%%%%%%%%%%%%%%%%%%%%%%%%%% Massive graviton effect
  %\cite{Blake:2013owa}
\bibitem{Blake:2013owa}
  M.~Blake, D.~Tong and D.~Vegh,
 % ``Holographic Lattices Give the Graviton an Effective Mass,''
  Phys.\ Rev.\ Lett.\  {\bf 112}, no. 7, 071602 (2014)
  %doi:10.1103/PhysRevLett.112.071602
  [arXiv:1310.3832 [hep-th]].
  %%CITATION = doi:10.1103/PhysRevLett.112.071602;%%
  %70 citations counted in INSPIRE as of 14 Dec 2015

%\cite{Amoretti:2014zha}
\bibitem{Amoretti:2014zha}
  A.~Amoretti, A.~Braggio, N.~Maggiore, N.~Magnoli and D.~Musso,
  %``Thermo-electric transport in gauge/gravity models with momentum dissipation,''
  JHEP {\bf 1409}, 160 (2014)
  doi:10.1007/JHEP09(2014)160
  [arXiv:1406.4134 [hep-th]].
  %%CITATION = doi:10.1007/JHEP09(2014)160;%%
  %49 citations counted in INSPIRE as of 24 Nov 2016

%\cite{Hu:2015dnl}
\bibitem{Hu:2015dnl}
  Y.~P.~Hu, H.~F.~Li, H.~B.~Zeng and H.~Q.~Zhang,
  %``Holographic Josephson Junction from Massive Gravity,''
  Phys.\ Rev.\ D {\bf 93}, no. 10, 104009 (2016)
  %doi:10.1103/PhysRevD.93.104009
  [arXiv:1512.07035 [hep-th]].
  %%CITATION = doi:10.1103/PhysRevD.93.104009;%%
  %4 citations counted in INSPIRE as of 25 Oct 2016




  %\cite{Cai:2014znn}
\bibitem{Cai:2014znn}
  R.~G.~Cai, Y.~P.~Hu, Q.~Y.~Pan and Y.~L.~Zhang,
  %``Thermodynamics of Black Holes in Massive Gravity,''
  Phys.\ Rev.\ D {\bf 91}, no. 2, 024032 (2015)
  %doi:10.1103/PhysRevD.91.024032
  [arXiv:1409.2369 [hep-th]].
  %%CITATION = doi:10.1103/PhysRevD.91.024032;%%
  %14 citations counted in INSPIRE as of 19 Nov 2015


%%%%%%%%%%%%%%%%%%%%%%%%%%%%%%%%%%%%%%%%%%%%%%%%%%%%%%%%%%%%%%%%slolutions of black hole in massive gravity
%\cite{Hendi:2015pda}
\bibitem{Hendi:2015pda}
  S.~H.~Hendi, S.~Panahiyan and B.~Eslam Panah,
  %``Charged Black Hole Solutions in Gauss-Bonnet-Massive Gravity,''
  JHEP {\bf 1601}, 129 (2016)
  %doi:10.1007/JHEP01(2016)129
  [arXiv:1507.06563 [hep-th]];
  %%CITATION = doi:10.1007/JHEP01(2016)129;%%
  %18 citations counted in INSPIRE as of 06 Nov 2016
%\cite{Hendi:2016pvx}
%\bibitem{Hendi:2016pvx}
  S.~H.~Hendi, B.~Eslam Panah and S.~Panahiyan,
  %``Massive charged BTZ black holes in asymptotically (a)dS spacetimes,''
  JHEP {\bf 1605}, 029 (2016)
  %doi:10.1007/JHEP05(2016)029
  [arXiv:1604.00370 [hep-th]];
  %%CITATION = doi:10.1007/JHEP05(2016)029;%%
  %6 citations counted in INSPIRE as of 06 Nov 2016
%\cite{Hendi:2016uni}
%\bibitem{Hendi:2016uni}
  S.~H.~Hendi, N.~Riazi and S.~Panahiyan,
  %``Holographical aspects of dyonic black holes: Massive gravity generalization,''
  arXiv:1610.01505 [hep-th].
  %%CITATION = ARXIV:1610.01505;%%
  %1 citations counted in INSPIRE as of 06 Nov 2016
%\cite{Hendi:2015hoa}
\bibitem{Hendi:2015hoa}
  S.~H.~Hendi, B.~E.~Panah and S.~Panahiyan,
  %``Einstein-Born-Infeld-Massive Gravity: adS-Black Hole Solutions and their Thermodynamical properties,''
  JHEP {\bf 1511}, 157 (2015)
  %doi:10.1007/JHEP11(2015)157
  [arXiv:1508.01311 [hep-th]];
    S.~H.~Hendi, B.~E.~Panah and S.~Panahiyan,
  %``Thermodynamical Structure of AdS Black Holes in Massive Gravity with Stringy Gauge-Gravity Corrections,''
  arXiv:1510.00108 [hep-th].
  %%CITATION = ARXIV:1510.00108;%%


%\cite{Hu:2016mym}
\bibitem{Hu:2016mym}
  Y.~P.~Hu, X.~X.~Zeng and H.~Q.~Zhang,
  %``Holographic Thermalization and Generalized Vaidya-AdS Solutions in Massive Gravity,''
  Phys.\ Lett.\ B {\bf 765}, 120 (2017)
  %doi:10.1016/j.physletb.2016.12.028
  [arXiv:1611.00677 [hep-th]].
  %%CITATION = doi:10.1016/j.physletb.2016.12.028;%%
  %2 citations counted in INSPIRE as of 18 Mar 2017




%%%%%%%%%%%%%%%%%%%%%%%%%%%%%%%%%%%%%%%%%%%%%%%%%%%%%%%%%%%%%%reviews on black holes in massive gravity
%\cite{Volkov:2013roa,Tasinato:2013rza,Babichev:2015xha}


%\cite{Volkov:2013roa}
\bibitem{Volkov:2013roa}
  M.~S.~Volkov,
  %``Self-accelerating cosmologies and hairy black holes in ghost-free bigravity and massive gravity,''
  Class.\ Quant.\ Grav.\  {\bf 30}, 184009 (2013)
  %doi:10.1088/0264-9381/30/18/184009
  [arXiv:1304.0238 [hep-th]].
  %%CITATION = doi:10.1088/0264-9381/30/18/184009;%%
  %65 citations counted in INSPIRE as of 18 Mar 2017

%\cite{Tasinato:2013rza}
\bibitem{Tasinato:2013rza}
  G.~Tasinato, K.~Koyama and G.~Niz,
  %``Exact Solutions in Massive Gravity,''
  Class.\ Quant.\ Grav.\  {\bf 30}, 184002 (2013)
  %doi:10.1088/0264-9381/30/18/184002
  [arXiv:1304.0601 [hep-th]].
  %%CITATION = doi:10.1088/0264-9381/30/18/184002;%%
  %44 citations counted in INSPIRE as of 18 Mar 2017

%\cite{Babichev:2015xha}
\bibitem{Babichev:2015xha}
  E.~Babichev and R.~Brito,
  %``Black holes in massive gravity,''
  Class.\ Quant.\ Grav.\  {\bf 32}, 154001 (2015)
  %doi:10.1088/0264-9381/32/15/154001
  [arXiv:1503.07529 [gr-qc]].
  %%CITATION = doi:10.1088/0264-9381/32/15/154001;%%
  %30 citations counted in INSPIRE as of 18 Mar 2017


%\cite{Xu:2015rfa}
\bibitem{Xu:2015rfa}
  J.~Xu, L.~M.~Cao and Y.~P.~Hu,
  %``P-V criticality in the extended phase space of black holes in massive gravity,''
  Phys.\ Rev.\ D {\bf 91}, no. 12, 124033 (2015)
  %doi:10.1103/PhysRevD.91.124033
  [arXiv:1506.03578 [gr-qc]].
  %%CITATION = doi:10.1103/PhysRevD.91.124033;%%
  %9 citations counted in INSPIRE as of 08 Dec 2015




%\cite{Cao:2015cza}
\bibitem{Cao:2015cza}
  L.~M.~Cao and Y.~Peng,
  %``Counterterms in Massive Gravity Theory,''
  arXiv:1509.08738 [hep-th];
  %%CITATION = ARXIV:1509.08738;%%
  %1 citations counted in INSPIRE as of 08 Dec 2015
%\cite{Cao:2015cti}
%\bibitem{Cao:2015cti}
  L.~M.~Cao, Y.~Peng and Y.~L.~Zhang,
  %``On dRGT massive gravity with degenerate reference metrics,''
  arXiv:1511.04967 [hep-th].
  %%CITATION = ARXIV:1511.04967;%%



%\cite{Davison:2013jba}
\bibitem{Davison:2013jba}
  R.~A.~Davison,
  %``Momentum relaxation in holographic massive gravity,''
  Phys.\ Rev.\ D {\bf 88}, 086003 (2013)
  %doi:10.1103/PhysRevD.88.086003
  [arXiv:1306.5792 [hep-th]].
  %%CITATION = doi:10.1103/PhysRevD.88.086003;%%
  %97 citations counted in INSPIRE as of 27 Oct 2016

%\cite{Blake:2013bqa}
\bibitem{Blake:2013bqa}
  M.~Blake and D.~Tong,
  %``Universal Resistivity from Holographic Massive Gravity,''
  Phys.\ Rev.\ D {\bf 88}, no. 10, 106004 (2013)
  %doi:10.1103/PhysRevD.88.106004
  [arXiv:1308.4970 [hep-th]].
  %%CITATION = doi:10.1103/PhysRevD.88.106004;%%
  %103 citations counted in INSPIRE as of 27 Oct 2016


%\cite{Davison:2013txa}
\bibitem{Davison:2013txa}
  R.~A.~Davison, K.~Schalm and J.~Zaanen,
%``Holographic duality and the resistivity of strange metals,''
  Phys.\ Rev.\ B {\bf 89}, 245116 (2014)
  [arXiv:1311.2451 [hep-th]].
  %%CITATION = ARXIV:1311.2451;%%
  %17 citations counted in INSPIRE as of 08 Sep 2014

%\cite{Adams:2014vza}
\bibitem{Adams:2014vza}
  A.~Adams, D.~A.~Roberts and O.~Saremi,
  %``The Hawking-Page Transition in Holographic Massive Gravity,''
  arXiv:1408.6560 [hep-th].
  %%CITATION = ARXIV:1408.6560;%%
  %3 citations counted in INSPIRE as of 30 Jan 2015

%\cite{Do:2016abo}
\bibitem{Do:2016abo}
  T.~Q.~Do,
  %``Higher dimensional nonlinear massive gravity,''
  Phys.\ Rev.\ D {\bf 93}, no. 10, 104003 (2016)
  %doi:10.1103/PhysRevD.93.104003
  [arXiv:1602.05672 [gr-qc]];
  %%CITATION = doi:10.1103/PhysRevD.93.104003;%%
  %2 citations counted in INSPIRE as of 06 Nov 2016
%\cite{Do:2016uef}
%\bibitem{Do:2016uef}
  T.~Q.~Do,
  %``Higher dimensional massive bigravity,''
  Phys.\ Rev.\ D {\bf 94}, no. 4, 044022 (2016)
  %doi:10.1103/PhysRevD.94.044022
  [arXiv:1604.07568 [gr-qc]].
  %%CITATION = doi:10.1103/PhysRevD.94.044022;%%
  %1 citations counted in INSPIRE as of 06 Nov 2016

\bibitem{vai}
%\cite{Vaidya:1951zz}
%\bibitem{Vaidya:1951zz}
  P.~C.Vaidya,
  %``The Gravitational Field of a Radiating Star,''
  Proc.\ Natl.\ Inst.\ Sci.\ India A {\bf 33}, 264 (1951);
  %213 citations counted in INSPIRE as of 29 Mar 2017
  %doi:10.1007/BF03173260.
  R. W. Lindquist, R. A. Schwartz, and C. W. Misner, Phys. Rev. \textbf{137}, B1364 (1965).

%\bibitem{vai}
  %P. C. Vaidya, Proc. Indian Acad. Sci. \textbf{A33}, 264 (1951);
  %R. W. Lindquist, R. A. Schwartz, and C. W. Misner, Phys. Rev. \textbf{137}, B1364 (1965).

%P. C. Vaidya, Proc. Indian Acad. Sci. (Math. Sci.) 33, 264(1951). doi:10.1007/BF03173260

%\cite{Dominguez:2005rt}
\bibitem{Dominguez:2005rt}
  A.~E.~Dominguez and E.~Gallo,
  %``Radiating black hole solutions in Einstein-Gauss-Bonnet gravity,''
  Phys.\ Rev.\ D {\bf 73}, 064018 (2006)
  doi:10.1103/PhysRevD.73.064018
  [gr-qc/0512150].
  %%CITATION = doi:10.1103/PhysRevD.73.064018;%%
  %22 citations counted in INSPIRE as of 23 Nov 2016

\bibitem{CCHK}
  R.~G.~Cai, L.~M.~Cao, Y.~P.~Hu and S.~P.~Kim,
  %``Generalized Vaidya Spacetime in Lovelock Gravity and Thermodynamics on
  %Apparent Horizon,''
  Phys.\ Rev.\  D {\bf 78}, 124012 (2008)
  [arXiv:0810.2610 [hep-th]].
  %%CITATION = PHRVA,D78,124012;%%
%%%%%%%%%%%%%%%%%%%%%%%%%%%%%%%%%%%%%%%%%%%%%%%%%%%%%%%%%%%%%%%%%%%%%%%%%%%%%

%%%%%%%%%%%%%%%%%%%%%%%%%%%%%%%%%%%%%%%%%%%%%%%%%%%%%%%%%%%%%%%%%%%%%%%%%%%%Unified first law and Misner-Sharp energy
\bibitem{Hayward} S.~A.~Hayward, %`` General laws of black-hole dynamics%
Phys.\ Rev.\ D \textbf{49}, 6467 (1994);
%\cite{Hayward:1993ph}
%\bibitem{Hayward:1993ph}
  S.~A.~Hayward,
  %``Quasilocal gravitational energy,''
  Phys.\ Rev.\ D {\bf 49}, 831 (1994)
  doi:10.1103/PhysRevD.49.831
  [gr-qc/9303030];
  %%CITATION = doi:10.1103/PhysRevD.49.831;%%
  %161 citations counted in INSPIRE as of 24 Nov 2016
%\bibitem{Hayward1}
S.~A.~Hayward,
%``Gravitational energy in spherical symmetry,''
Phys.\ Rev.\ D \textbf{53}, 1938 (1996) [arXiv:gr-qc/9408002];
%\bibitem{Hayward2}
S.~A.~Hayward,
%``Unified first law of black-hole dynamics and relativistic thermodynamics,''
Class.\ Quant.\ Grav.\ \textbf{15}, 3147 (1998) [arXiv:gr-qc/9710089].


%%%%%%%%%%%%%%%%%%%%%%%%%%%%%%%%%%%%%%%%%%%%%%%%%%%%%%%%%%%%%%%%%%%%%%%%%%%%%%%%%%%%



%%%%%%%%%%%%%%%%%%%%%%%%%%%%%%%%%%%%%%%%%%%%%%%%%%%%%%%%%%%%%%%%%%Misner-Sharp mass
  \bibitem{mae}
  H.~Maeda and M.~Nozawa,
  %``Generalized Misner-Sharp quasi-local mass in Einstein-Gauss-Bonnet
  %gravity,''
  Phys.\ Rev.\  D {\bf 77}, 064031 (2008)
  [arXiv:0709.1199 [hep-th]].


%\cite{Cai:2009qf}

  %%CITATION = doi:10.1103/PhysRevD.90.024062;%%
  %9 citations counted in INSPIRE as of 27 Oct 2016




%%%%%%%%%%%%%%%%%%%%%%%%%%%%%%%%%%%%%%%%%%%%%%%%%%%%%%%%%%%%%%%%%%%%%%%%%%%%%%%Thermodynamics on the apparent horizon
%\cite{Cai:2005ra}
\bibitem{Cai-Kim}
  R.~G.~Cai and S.~P.~Kim,
  %``First law of thermodynamics and Friedmann equations of
  %Friedmann-Robertson-Walker universe,''
  JHEP {\bf 0502}, 050 (2005)
  [arXiv:hep-th/0501055].
  %%CITATION = JHEPA,0502,050;%%

%\cite{Cai:2008ys}
\bibitem{Cai:2008ys}
  R.~G.~Cai, L.~M.~Cao and Y.~P.~Hu,
  %``Corrected Entropy-Area Relation and Modified Friedmann Equations,''
  JHEP {\bf 0808}, 090 (2008)
  [arXiv:0807.1232 [hep-th]].
  %%CITATION = ARXIV:0807.1232;%%
  %98 citations counted in INSPIRE as of 30 Jan 2015
\bibitem{self2}
%\cite{Zhang:2015bva}
%\bibitem{Zhang:2015bva}
  H.~Zhang,
  %``Solutions in Gravity Theories Emerge from Thermodynamics : A Mini Review of a Recent Progress in Graviti-Thermodynamics,''
  The Universe {\bf 3}, no. 1, 30 (2015).
  %2 citations counted in INSPIRE as of 29 Mar 2017

   \bibitem{self31}
  H. Zhang, S. Hayward, X. H. Zhai, and X.Z. Li, Phys. Rev. D89(2014)064052; H.~Zhang and X.~Z.~Li,
  %``From thermodynamics to the solutions in gravity theory,''
  Phys.\ Lett.\ B {\bf 737} (2014) 395
  [arXiv:1406.1553 [gr-qc]]; Hongsheng Zhang, Dao-Jun Liu, and Xin-Zhou Li, Phys. Rev. D 90, 124051 (2014)[arxiv: 1405.7530];  D.~He and Q.~y.~Cai,
  %``Gravitational correlation, black hole entropy and information conservation,''
  arXiv:1609.05825 [hep-th]; H.~W.~Tan, J.~B.~Yang, T.~M.~He and J.~Y.~Zhang,
  %``A modified thermodynamics method to generate exact solutions of the Einstein equations,''
  arXiv:1609.04181 [gr-qc].

%%%%%%%%%%%%%%%%%%%%%%%%%%%%%%%%%%%%%%%%%%%%%%%%%%%%%%%%%%%%%%%%%%%%%%%%%%%%%%%
%\bibitem{uni} repeat
 %S. A. Hayward, Phys. Rev. D{\bf49}, 831(1994) [arXiv:gr-qc/9303030];Phys. Rev. D{\bf53}, 1938(1996) [arXiv:gr-qc/9408002];
  %``Unified first law of black hole dynamics and relativistic thermodynamics,''
  %Class.\ Quant.\ Grav.\  {\bf 15}, 3147(1998)[gr-qc/9710089].
%%%%%%%%%%%%%%%%%%%%%%%%%%%%%%%%%%%%%%%%%%%%%%%%%%%%%%%%%%%%%%%%%%%%%%%%%%%%%%%

\bibitem{Cai:2009qf}
  R.~G.~Cai, L.~M.~Cao, Y.~P.~Hu and N.~Ohta,
  %``Generalized Misner-Sharp Energy in f(R) Gravity,''
  Phys.\ Rev.\ D {\bf 80}, 104016 (2009)
  %doi:10.1103/PhysRevD.80.104016
  [arXiv:0910.2387 [hep-th]];
  %%CITATION = doi:10.1103/PhysRevD.80.104016;%%
  %50 citations counted in INSPIRE as of 27 Oct 2016

%\cite{Zhang:2014goa}
\bibitem{Zhang:2014goa}
  H.~Zhang, Y.~Hu and X.~Z.~Li,
  %``Misner-Sharp Mass in $N$-dimensional $f(R)$ Gravity,''
  Phys.\ Rev.\ D {\bf 90}, no. 2, 024062 (2014)
  %doi:10.1103/PhysRevD.90.024062
  [arXiv:1406.0577 [gr-qc]].
%%%%%%%%%%%%%%%%%%%%%%%%%%%%%%%%%%%%%%%%%%%%%%%%%%%%%%%%%%%%%%%%%%%%%%%%%%%%%%%%%%Thermodynamics in f(R) gravity


\bibitem{aka} M.~Akbar and R.~G.~Cai,
  %``Thermodynamic Behavior of Friedmann Equation at Apparent Horizon of FRW
  %Universe,''
  Phys.\ Rev.\  D {\bf 75}, 084003 (2007)
  [arXiv:hep-th/0609128];
  %%CITATION = PHRVA,D75,084003;%%
   M.~Akbar and R.~G.~Cai,
  %``Friedmann equations of FRW universe in scalar-tensor gravity, f(R)  gravity
  %and first law of thermodynamics,''
  Phys.\ Lett.\  B {\bf 635}, 7 (2006)
  [arXiv:hep-th/0602156];
  %%CITATION = PHLTA,B635,7;%%
  M.~Akbar and R.~G.~Cai,
  %``Thermodynamic Behavior of Field Equations for f(R) Gravity,''
  Phys.\ Lett.\  B {\bf 648}, 243 (2007)
  [arXiv:gr-qc/0612089].
  %%CITATION = PHLTA,B648,243;%%


%\cite{Eling:2006aw}
\bibitem{Eling:2006aw}
  C.~Eling, R.~Guedens and T.~Jacobson,
  %``Non-equilibrium thermodynamics of spacetime,''
  Phys.\ Rev.\ Lett.\  {\bf 96}, 121301 (2006)
  %doi:10.1103/PhysRevLett.96.121301
  [gr-qc/0602001];
  %%CITATION = doi:10.1103/PhysRevLett.96.121301;%%
  %224 citations counted in INSPIRE as of 27 Oct 2016
%\cite{Jacobson:1995ab}
%\bibitem{Jacobson:1995ab}
  T.~Jacobson,
  %``Thermodynamics of space-time: The Einstein equation of state,''
  Phys.\ Rev.\ Lett.\  {\bf 75}, 1260 (1995)
  %doi:10.1103/PhysRevLett.75.1260
  [gr-qc/9504004].
  %%CITATION = doi:10.1103/PhysRevLett.75.1260;%%
  %990 citations counted in INSPIRE as of 27 Oct 2016

%\cite{Wang:1998qx}
\bibitem{Wang:1998qx}
  A.~Wang and Y.~Wu,
  %``Generalized Vaidya solutions,''
  Gen.\ Rel.\ Grav.\  {\bf 31}, 107 (1999)
  %doi:10.1023/A:1018819521971
  [gr-qc/9803038].
  %%CITATION = doi:10.1023/A:1018819521971;%%
  %96 citations counted in INSPIRE as of 18 Mar 2017



 \bibitem{LLX}
  P.~Li, X.~z.~Li and P.~Xi,
  %``Black hole solutions in de Rham-Gabadadze-Tolley massive gravity,''
  Phys.\ Rev.\ D {\bf 93} (2016) no.6,  064040
  doi:10.1103/PhysRevD.93.064040
  [arXiv:1603.06039 [gr-qc]]; P.~Li, X.~z.~Li and P.~Xi,
  %``Analytical expression for a class of spherically symmetric solutions in Lorentz breaking massive gravity,''
  Class.\ Quant.\ Grav.\  {\bf 33}, no. 11, 115004 (2016)
  doi:10.1088/0264-9381/33/11/115004
  [arXiv:1503.08952 [gr-qc]].




%\cite{Babichev:2014tfa}
\bibitem{Babichev:2014tfa}
  E.~Babichev and A.~Fabbri,
  %``Rotating black holes in massive gravity,''
  Phys.\ Rev.\ D {\bf 90}, 084019 (2014)
  doi:10.1103/PhysRevD.90.084019
  [arXiv:1406.6096 [gr-qc]].
  %%CITATION = doi:10.1103/PhysRevD.90.084019;%%
  %17 citations counted in INSPIRE as of 21 Nov 2016





%%%%%%%%%%%%%%%%%%%%%%%%%%%%%%%%%%%%%%%%%%%black branes in large D
%\cite{Emparan:2015gva}
\bibitem{Emparan:2015gva}
  R.~Emparan, R.~Suzuki and K.~Tanabe,
  %``Evolution and End Point of the Black String Instability: Large D Solution,''
  Phys.\ Rev.\ Lett.\  {\bf 115}, no. 9, 091102 (2015)
  doi:10.1103/PhysRevLett.115.091102
  [arXiv:1506.06772 [hep-th]];
  %%CITATION = doi:10.1103/PhysRevLett.115.091102;%%
  %13 citations counted in INSPIRE as of 21 Nov 2016
%\cite{Emparan:2016sjk}
%\bibitem{Emparan:2016sjk}
  R.~Emparan, K.~Izumi, R.~Luna, R.~Suzuki and K.~Tanabe,
  %``Hydro-elastic Complementarity in Black Branes at large D,''
  JHEP {\bf 1606}, 117 (2016)
  doi:10.1007/JHEP06(2016)117
  [arXiv:1602.05752 [hep-th]].
  %%CITATION = doi:10.1007/JHEP06(2016)117;%%
  %7 citations counted in INSPIRE as of 21 Nov 2016


%%%%%%%%%%%%%%%%%%%%%%%%%%%%%%%%%%%%%%%%%%%%%%%%%%%%%%%%%%%%%Holographic Thermalization

%\bibitem{Balasubramanian1} V. Balasubramanian {\it et al.}, Phys. Rev. Lett. 106, 191601 (2011).
\bibitem{Balasubramanian1}
%\cite{Balasubramanian:2010ce}
%\bibitem{Balasubramanian:2010ce}
  V.~Balasubramanian {\it et al.},
  %``Thermalization of Strongly Coupled Field Theories,''
  Phys.\ Rev.\ Lett.\  {\bf 106}, 191601 (2011)
  %doi:10.1103/PhysRevLett.106.191601
  [arXiv:1012.4753 [hep-th]].
  %%CITATION = doi:10.1103/PhysRevLett.106.191601;%%
  %204 citations counted in INSPIRE as of 29 Mar 2017


%\bibitem{Balasubramanian2} V. Balasubramanian {\it et al.},Phys. Rev. D 84, 026010 (2011); V. Balasubramanian and S. F. Ross,  Phys. Rev. D 61, 044007 (2000).
%\bibitem{Balasubramanian61}

\bibitem{Balasubramanian2}
%\cite{Balasubramanian:2011ur}
%\bibitem{Balasubramanian:2011ur}
  V.~Balasubramanian {\it et al.},
  %``Holographic Thermalization,''
  Phys.\ Rev.\ D {\bf 84}, 026010 (2011)
  %doi:10.1103/PhysRevD.84.026010
  [arXiv:1103.2683 [hep-th]];
  %%CITATION = doi:10.1103/PhysRevD.84.026010;%%
  %211 citations counted in INSPIRE as of 29 Mar 2017
%\cite{Balasubramanian:1999zv}
%\bibitem{Balasubramanian:1999zv}
  V.~Balasubramanian and S.~F.~Ross,
  %``Holographic particle detection,''
  Phys.\ Rev.\ D {\bf 61}, 044007 (2000)
  %doi:10.1103/PhysRevD.61.044007
  [hep-th/9906226].
  %%CITATION = doi:10.1103/PhysRevD.61.044007;%%
  %143 citations counted in INSPIRE as of 29 Mar 2017

%\cite{Hu:2016mym}

  %%CITATION = ARXIV:1611.00677;%%



  %%CITATION = PHRVA,D62,124023;%%







  %%CITATION = PHRVA,D62,124023;%%








\end{thebibliography}
\end{document}